# Enhancing extraordinary transmission of light through a metallic nano slit with a nano cavity antenna

Yanxia Cui[1] and Sailing He[1, 2, *]
[1]Centre for Optical and Electromagnetic Research, and Joint Research Centre of Photonics of the Royal Institute of Technology (Sweden) and Zhejiang University, Zijingang Campus, Zhejiang University, China
[2] Division of Electromagnetic Theory, School of Electrical Engineering, Royal Institute of Technology, S-100 44 Stockholm, Sweden
*Corresponding author: sailing@kth.se

The extraordinary transmission of light through a nano slit in a metal film is enhanced by introducing a nano cavity antenna formed by a nearby metallic nano-strip over the slit opening. For a fixed wavelength, the width of the metallic nano-strip should be chosen to make the horizontal metal-insulator-metal waveguide of finite length resonant as a Fabry-Perot cavity. When such a cavity antenna is used to enhance the transmission through a non-resonant nano slit, the slit should be opened at a position with maximal magnetic field in the horizontal resonant cavity. It is shown that an optimized cavity antenna can enhance greatly the transmission of light through a non-resonant nano slit (by about 20 times) or a resonant nano slit (by 124%). The transmission spectrum of the nano slit can also be tuned by adjusting the width of the metallic nano-strip. Such a transmission enhancement with a nano cavity antenna is studied for the first time and the physical mechanism is explained.

Extraordinary transmission of light through nano holes, including single apertures and aperture arrays, has attracted considerable attention and been widely investigated in the recent years [1-7]. Surface plasmons polaritons (SPPs) [8] play an essential role in this phenomenon, and various applications have been made, such as enhanced spectroscopy and single molecule fluorescence [9] (for biological applications), photolithography, light-emitting diode (LED) design and ultrafast photodetectors [10].

In the case of single nano slit (or hole), the transmission can be enhanced by some periodic corrugations surrounding the slit [2, 6, 7]. In this letter we propose a different approach to the enhancement of extraordinary transmission through the single nano slit, namely, forming a cavity antenna by introducing a nearby metallic nano-strip over the slit opening (see FIG. 1). Intuitively, the metallic nano-strip over the nano slit seems to block the incident light. However, through the formation of a resonant nano cavity antenna, the metallic nano-strip can assist to couple more incident light into the nano slit and thus enhance the transmission. Note that optical nano antennas for other applications have attracted much attention very recently (see [11-13] in the recent special issue of Nature Photonics) while the present receiving nano antenna has a specific physical mechanism of resonant cavity antenna for the enhancement of extraordinary transmission. Such a transmission enhancement with a nano cavity antenna is studied for the first time in the present letter.

FIG. 1 shows the two-dimensional (2D) schematic diagram for the present nano cavity antenna formed by a metallic rectangular nano-strip (with thickness $H_p$ and width $W_p$) over (with a distance of $d$) a metallic nano slit of width $W_s$ in a metal film of thickness $t$. The top interface of the metal film is at $z=0$, and the nano slit is centered at $x=0$. A plane wave of TM polarization (the magnetic field is perpendicular to the $x$-$z$ plane) with wavelength $\lambda_0=1\mu m$ impinges normally on the top of the structure. Both the film and the nano-strip are made of silver, whose permittivity is obtained from Ref. [5]. All the results are calculated with a 2D finite element method (FEM). The transmission efficiency $\eta$ is defined as the ratio of the integration of the $z$-component of the Poynting vector over the output opening (with or without the nano-strip) to that over the input opening of the bare slit (without the nano-strip).

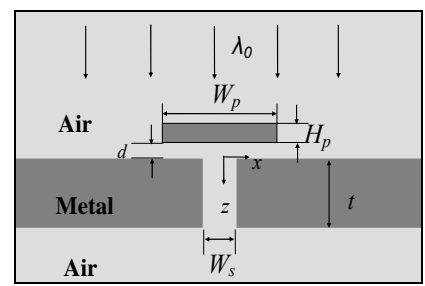

FIG. 1. Configuration of the nano cavity antenna formed by a nearby metallic nano-strip over the opening of a nano slit milled in a metallic thin film.

A nano slit milled in a metallic thin film can form a vertical metal-insulator-metal (MIM) waveguide of finite length, which can give Fabry-Perot (F-P) resonance in the narrow slit region [14, 15]. The transmission is maximal due to constructive interference for a length phase of even integer of $\pi/2$ (related to the effective wavelength in the MIM waveguide) and is minimal for a length phase of odd integer of $\pi/2$ (due to destructive interference) [5]. A horizontal MIM waveguide cavity can also be formed by putting a metallic nano-strip over a metal film with an air gap in-between. FIG. 2(a) shows the distribution of the magnetic field modulus ($|H_y|$) when we put a silver nano-strip (with $W_p=1\mu m$ and $H_p=300nm$) over (with $d=44nm$) the silver film to form constructive interference of three peaks in the horizontal F-P cavity (close to a $3\pi$ phase length waveguide; due to some spreading at the two ends the light confinement region is a bit longer than the width of the metallic nano-strip). Note that the effective index of the MIM waveguide is larger than 1 (for air) due to the enhancement of the refractive index [15]. Also note that the resonant mode loses the up-down symmetry as the bottom air-metal interface is infinitely long. One sees obviously that the metallic nano-strip collects the incident light effectively due to the resonance in the cavity below the nano-strip, and localized field of high magnitude is accumulated within the horizontal cavity. Then we make a nano slit in the metal film at a position where the magnetic field in the horizontal resonant cavity is maximal and wish to enhance the extraordinary transmission through the nano slit in the metal film.

The influence of the nano slit configuration to the field distribution in the horizontal cavity is different depending on whether the vertical nano slit is in resonance or not. First we consider the case of a non-resonant nano slit with destructive interference. The inset of FIG. 2(b) shows the $|H_y|$ distribution for a bare nano slit with opening width $W_s=25nm$ (all the nano slits studied in this letter is set with this width) milled in a silver film of 390 $nm$ thickness (as a vertical waveguide of $3\pi/2$ phase length). In order to show the details more clearly, we plot the field distribution with doubled magnitude in this inset. The transmission efficiency for this bare nano slit is very low ($\eta = 0.6$) and the nano slit causes little distortion of the field reflected by the silver film. FIG. 2(b) shows the field distribution when there is a silver nano-strip over this non-resonant nano slit (opened at the center where the magnetic field is maximal in the horizontal resonant cavity). Compared with FIG. 2(a), we find that the mode pattern in the horizontal cavity keeps almost the same, which means that the influence of the non-resonant nano slit to the mode distribution in the horizontal cavity is quite small. On the other hand, the resonant cavity antenna over the non-resonant nano slit brings an enhanced transmission, and the energy flowing out of the nano slit aperture is enhanced by a factor of 5 as compared with that for the non-resonant bare nano slit in the inset of FIG. 2(b). We then move the nano slit along the $x$ direction to another position of maximal magnetic field in the horizontal cavity, as shown in FIG. 2(c). Comparing FIG. 2(b) and (c), one does not see any obvious change in the distribution of the field modulus in the horizontal cavity or the vertical nano slit, and $\eta$ remains the same. However, when we move the nano slit opening to a position where the magnetic field is minimal in the horizontal resonant cavity [as shown in FIG. 2(d)], the collected light by the resonant cavity can hardly pass through the nano slit and $\eta$ is nearly zero. Note that inside the horizontal resonant cavity the electric field component $|E_z|$ is maximal (minimal) at a position where the magnetic field is minimal (maximal). The above position-dependence of the enhanced transmission can be explained as follows. Larger magnetic field generates around the slit opening larger surface current ($\vec{J} = \nabla \times \vec{H}$), which would give larger SPPs into the non-resonant nano slit. Thus we can conclude that the present cavity antenna can enhance efficiently the optical transmission through the non-resonant nano slit opened at a position with maximal magnetic field in the horizontal resonant cavity.

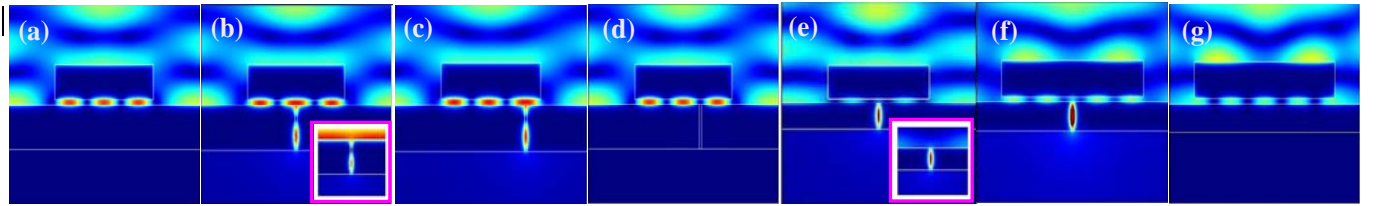

FIG. 2. (Color online) Distributions of magnetic filed $|H_y|$ for (a) a horizontal resonant MIM cavity ($W_p=1\mu m$); (b-d) a horizontal resonant cavity over a non-resonant nano slit opened at different positions; (e) the same horizontal cavity over a resonant nano slit; (f) a longer horizontal cavity ($W_p=1.46\mu m$) over a resonant nano slit; and (g) the same long horizontal cavity with the resonant nano slit removed. The insets in (b) and (e) show the $|H_y|$ distributions for the corresponding cases of a bare nano slit.

The inset of FIG. 2(e) shows the $|H_y|$ distributions for a resonant bare nano slit with $t=240nm$ (corresponding to a vertical MIM waveguide of $\pi$ phase length). An output transmission efficiency $\eta=7.8$ is achieved, attributed to the constructive interference in the vertical MIM waveguide of finite length. Meanwhile, we observe a strongly distorted reflection pattern with small field right above the slit opening. When the silver nano-strip (same as FIG. 2(a-d)) is put over the resonant nano slit as shown in FIG. 2(e), the mode distribution in the horizontal cavity is distorted significantly and quite different from that in FIG. 2(a). When the nano slit is in resonance, the occurrence of small magnetic field near the slit opening should be mainly determined by the magnetic field of the resonant mode of the vertical nano slit rather than the interference field of the horizontal resonant cavity. Consequently, the horizontal cavity can no longer keep its optimal internal field distribution for the harvest of light and the cavity antenna does not work any more. Thus the energy flowing into the nano slit is small and $\eta$ is only 3.6, lower than $\eta$ in the inset of FIG. 2(e). The strip width can be adjusted to improve the transmission. FIG. 2(f) shows the $|H_y|$ distribution when the nano-strip width is adjusted to $W_p=1460nm$, and the corresponding transmission efficiency reaches $\eta=6.7$ (further enhancement can be achieved by optimizing the height of the nano-strip to $H_p=50nm$; see FIG. 4(a) below). For comparison, the corresponding $|H_y|$ distribution when the nano slit is removed as shown in FIG. 2(g). Again, one sees that the resonant nano slit will change the field pattern in the

horizontal cavity. We have also considered some other situations when the slit thickness is apart from the resonant case and found that the mode distortion in the horizontal cavity is relatively smaller as compared with the resonant case.

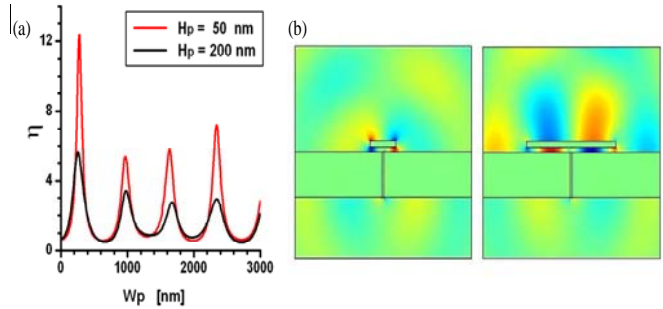

FIG. 3. (Color online) (a) Dependence of the transmission on the width of the metallic nano-strip for different nano-strip heights with $W_s$=25$nm$, $t$=390$nm$ and $d$=44$nm$; (b) Distributions of electric field $E_z$ for (left)$W_p$=280$nm$ (right) or 970$nm$ when $H_p$=50$nm$.

FIG. 3(a) shows the transmission as a function of the width of the horizontal cavity antenna ($W_p$) for different nano-strip heights [($H_p$=50$nm$ (red curve) or $H_p$=200$nm$ (black curve)] when the nano slit is non-resonant. One sees that the transmitted optical energy varies with a period around 700$nm$ (i.e., the effective propagating wavelength in the horizontal MIM waveguide). One also sees several interesting features from FIG. 3(a). First, the transmission peaks for $H_p$=50$nm$ are much higher than those for $H_p$=200$nm$. This can be explained as follows. A thick metallic nano-strip tends to block light passing through it. However, if $H_p$ deceases to be close to the skin depth (about 20$nm$) of the metal, light incident on the top surface of the metallic nano-strip can transmit through it and the down-propagating energy will serve as an additional source to the horizontal cavity, which makes a thinner silver nano-strip better for transmission enhancement. Secondly, in the bare nano slit case, due to the absorption by the slit's metallic walls, the envelope of the transmission curve drops gradually as the film thickness increases [5]. However, when there is a metallic nano-strip over the nano slit, the envelope of the oscillating curve in general does not decrease monotonously as the horizontal strip width $W_p$ increases [as one can see clearly from the case of $H_p$=50$nm$ in FIG. 3(a); note that the fourth peak is even a little bit higher than the third peak for the case of $H_p$=200$nm$ in FIG. 3(a)]. This feature should come from the competition between the metal loss and the light-collecting area of the cavity antenna. Since the top surface of the nano-strip serves as a converter to transform incident light into SPPs, a cavity antenna with a wider nano-strip has a larger collecting area for the incident energy. On the other hand, a wider metallic nano-strip also suffers more propagating loss of SPPs. In FIG. 3(a), one sees that the peak values (starting from the second one) of the transmission for $H_p$=50$nm$ increase gradually, which indicates that the increment of the collected light energy is larger than the dissipation loss as the nano-strip becomes wider. Finally, we notice in FIG. 3(a) that the first peak ($W_p$=280$nm$, $H_p$=50$nm$) with quite small collecting area is abnormally high with $\eta$=12.3 (about 20 times higher than $\eta$=0.6 for the corresponding non-resonant nano slit in the inset of FIG. 2(b); even higher than $\eta$=7.8 for a resonant bare nano slit). To explain this, we show the $E_z$ field distributions for the cases of the first and second peaks in FIG. 3(b). The charge profiles on the top surface of the nano-strip are determined by $E_z$ (i.e., the perpendicular component of the local $E$-field) [5]. When $W_p$ is about 280$nm$, surface charges with opposite signs accumulate strongly around the up corners of the nano-strip [see the left figure of FIG. 3(b)], which behaves as an efficient electric dipole. This electric dipole, which is very close to the slit opening (the dipole is right at the opening in Ref. [5]), may give additional contribution to the transmission enhancement. In other words, this narrow and thin nano-strip enhances the transmission through the nano-slit via its role as both an electric dipole and a cavity antenna (note that the magnetic field $|H_y|$ is still in its maximal at the center of this narrow horizontal cavity). For a higher order peak corresponding to a wider nano-strip, the equivalent electric dipole (associated with the accumulated charges at the up corners of the strip) becomes weaker [see the right figure of FIG. 3(b)] and farther away from the nano slit, and consequently their contribution to the transmission enhancement is much smaller and the mechanism of resonant cavity antenna dominates.

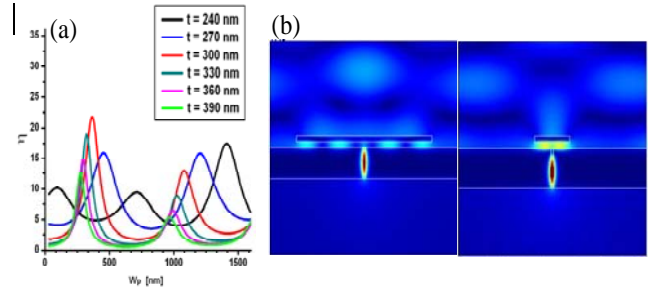

FIG. 4. (Color online) (a) Dependence of the transmission efficiency on the width of the metallic nano-strip for different nano slit thicknesses with $H_p$=50$nm$, $d$=44$nm$, $W_s$=25$nm$; (b) Distributions of magnetic field $|H_y|$ for (left) $t$=240$nm$ and $W_p$=1410$nm$ or (right) $t$=300$nm$ and $W_p$=360$nm$.

To optimize further the transmission through the nano slit assisted by this cavity antenna, we scan the transmission efficiency for different slit thicknesses (over an effective propagating wavelength in the horizontal MIM waveguide) as $W_p$ varies for a very thin silver nano-strip ($H_p$=50$nm$). The wavelength is set as $\lambda_0$=1$\mu m$, and the transmission results are shown in FIG. 4(a). The black curve (t=240nm) corresponds to a resonant nano slit while the other curves represent non-resonant nano slits. Different from FIG. 3(a), one sees in FIG. 4(a) that the position of the first peak of the black curve shifts to 90$nm$ and the transmission could hardly be greatly enhanced by the silver nano-strip of this narrow width due to the extremely small collecting area. As $W_p$ increases, through the competition between the collecting area and the metal loss, the transmission can be greatly enhanced to $\eta$=17.5 at $W_p$=1410$nm$ (124% larger than $\eta$=7.8 for a resonant bare nano slit). The left figure of FIG. 4(b) shows the corresponding distribution of $|H_y|$ $|H_y|$ for such an optimized cavity antenna for the enhancement of extraordinary transmission through a resonant nano slit. We also see in FIG. 4(a) that the largest transmission $\eta=22$ can be achieved with nano-strip width $W_p=360nm$ for a thicker film $t=300$ $nm$ (in red, a non-resonant nano slit). The corresponding distribution of $|H_y|$ is shown in the right figure of FIG. 4(b). Note that this largest transmission is 182% larger than the highest transmission ($\eta=7.8$) for a resonant bare nano slit. Thus, to achieve the highest transmission, we can replace the traditional resonant bare nano slit with a suitable non-resonant nano slit assisted with a resonant cavity antenna.

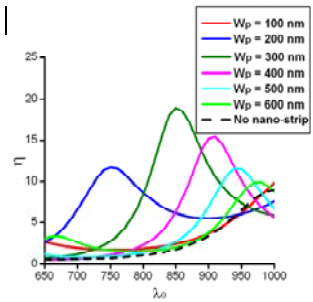

FIG. 5. (Color online) Transmission spectrum for various metallic nano-strip widths when $t=240nm$ (the dashed line is for the corresponding bare nano slit). $H_p=50nm$, $d=44nm$, $W_s=25nm$.

The horizontal cavity antenna will also change the transmission spectrum of the nano slit aperture. Simulations are carried out in the wavelength range of 650-1000$nm$, and the wavelength-dependent permittivity is described by a Drude plus two-diploes Lorentzian form [16]. In FIG. 5, we show the transmission spectrum for various antenna widths ($W_p$) when $H_p=50nm$ and $d=44nm$. The spectrum property for a bare nano slit ($W_s=25nm$, $t=240nm$) resonant at $\lambda_0=1.0\mu m$ is shown by the dashed line, from which one sees that the transmission efficiency is much lower when the nano slit is far from resonance. When a narrow silver nano-strip ($W_p=100nm$) is introduced, the curve (in red) is still approximate to that for the bare nano slit, indicating that the influence of a narrow silver nano-strip is weak. However, for a silver nano-strip with a large width, the influence becomes significant. When $W_p=200nm$ (in blue), a transmission peak appears around $\lambda_0=750nm$. As the width of the nano-strip increases, the transmission peak has a red shift and approaches the resonant wavelength of the bare nano slit.

In conclusion, we have introduced for the first time a resonant cavity antenna to enhance the transmission efficiency through some metallic nano slits. The physical mechanism has been explained in details. It has been demonstrated that the metallic nano-strip over the nano slit forms a horizontal cavity. To achieve an efficient receiving antenna for light conversion (from incident light to SPPs) and further transmission enhancement, the width of the metallic nano-strip should be carefully designed to form a resonant cavity and the height of the nano-strip should also be small enough. For a resonant nano slit, the transmission efficiency can be increased by 124% with a suitable cavity antenna. If this resonant nano slit can be replaced with a non-resonant nano slit (through changing the film thickness or wavelength), a suitable cavity antenna can enhance the transmission by 182% (compared with the extraordinary transmission through a resonant bare nano slit). We expect that this resonant nano cavity antenna can play a useful role in the research of nano-apertures and their applications.

This work has been supported partially by the National Natural Science Foundation of China (grants 60688401 and 60677047), the National Basic Research Program (973) of China (No. 2004CB719800), U. S. Air Force Office of Scientific Research (Asia Office), and the Swedish Research Council (VR) grant on metamaterial. The authors would also like to thank Dr. Yi Jin and Zhechao Wang for helpful discussions.